\documentclass{iaus}
\usepackage{graphicx}

\title[AO Observations of LBGs at $z\sim3$] %% give here short title %%
{Adaptive Optics Imaging of Lyman Break Galaxies as Progenitors of Spheroids in the Local Universe
 \thanks{Based on data collected at Subaru Telescope, which is operated by the
 National Astronomical Observatory of Japan.}}

\author[Akiyama et al.]   %% give here short author list %%
{M. Akiyama$^1$ Y.Minowa$^2$ N.Kobayashi$^3$ K.Ohta$^4$ \and I.Iwata$^5$}

\affiliation{
$^1$Subaru Telescope, National Astronomical Observatory of Japan,
650 North A'ohoku Place, Hilo, HI, 96720, USA \break email: akiyama@naoj.org\\[\affilskip]
$^2$Optical and Infrared Astronomy Division, NAOJ, 
Mitaka, 181-8588, Japan \\[\affilskip]
$^3$Institute of Astronomy, University of Tokyo,
Mitaka, 181-0015, Japan \\[\affilskip]
$^4$Department of Astronomy, Kyoto University,
Kyoto 606-8502, Japan \\[\affilskip]
$^5$Okayama Astrophysical Observatory, NAOJ,
Okayama 719-0232, Japan \\[\affilskip]
}

\pubyear{2007}
\volume{245}  %% insert here IAU Symposium No.
\pagerange{1-5}
\date{2007/08/31}
\jname{Formation and Evolution of Galaxy Bulges}
\editors{M.Bureau et al., eds.}
\begin{document}

\maketitle

\begin{abstract}
In order to reveal the stellar mass distribution of $z\sim3$ galaxies, we are conducting deep 
imaging observations of U-dropout Lyman Break Galaxies (LBGs) with Adaptive Optics (AO) 
systems in $K$-band, which corresponds to rest-frame $V$-band of $z\sim3$ galaxies. The results of 
the Subaru intensive-program observations with AO36/NGS/IRCS indicate that 1) the K-band peaks 
of some of the LBGs brighter than $K=22.0$ mag show significant offset from those in the optical 
images, 2) the $z\sim3$ $M_{V}^{*}$ LBGs and 
serendipitously observed Distant Red Galaxies (DRGs) have flat profiles similar to disk 
galaxies in the local universe (i.e., S\'ersic with $n<2$), and 3) the surface stellar mass 
densities of the $M_{V}^{*}$ LBGs are 3-6 times larger than those of disk galaxies at $z=0-1$. 
Considering the lack of $n>2$ systems among the luminous $z\sim3$ 
LBGs and DRGs, and their strong spatial clustering, we infer that the dense $n<2$ disk-like 
structures evolve into the $n>2$ spheroids of nearby galaxies through relaxations due to 
major merger events.
\end{abstract}

\firstsection % if your document starts with a section,
\section{Introduction}

The statistical evolution of the morphology of galaxies is one of the
most fundamental observational clues to understand the establishing 
process of the Hubble sequence of galaxies seen in the local universe.
Especially rest-frame optical morphology is important because it
reflects stellar mass distribution of galaxies. High-resolution imaging
observations with Hubble Space Telescope (HST), covering rest-frame
optical wavelength up to $z\sim1$, have shown that the Hubble
sequence is already established at $z\sim1$. Morphological studies
on galaxies at higher redshifts started following the discovery of
a large number of high redshift galaxies using Lyman Break selection (Steidel et al. 2003).
HST NICMOS observations of the Hubble Deep Field North (HDFN) show that
the morphology of the LBGs is essentially independent of the
wavelength up to the rest-frame 4000{\AA} (Giavalisco 2002).

However, the NICMOS observations are not sufficient to 
conclude the distribution of the stellar mass
in $z\sim3$ LBGs. The NICMOS observations covers up to the $H$-band,
which is just at 4000{\AA} in the rest-frame of $z\sim3$ galaxies. 
In addition, the NICMOS observations are mostly limited to the 
$z\sim3$ LBGs in the small HDFN area and only include object as
bright as $M_{B}\sim-22$mag, which is still 0.5mag fainter than 
the characteristic absolute magnitude of the $z\sim3$ LBGs ($M_{V}^{*}=-24.0$ mag).

In order to examine the rest-frame $V$-band morphology of
the $z\sim3$ LBGs covering a wide luminosity range ($M_{V}^{*}-0.5$ mag $-$
$M_{V}^{*}+3.0$ mag), we conducted AO-assisted imaging 
observations of 36 of them in the $K$-band, which corresponds to the rest-frame $V$-band
in $z\sim3$ galaxies (Akiyama et al. 2007). Thirty one of the LBGs are detected in the deep
imaging observations with typical effective integration time of 5 hours. 

\section{Results of AO imaging observations of the $z\sim3$ LBGs}

\begin{figure}[t]
\begin{center}
\includegraphics[width=110mm]{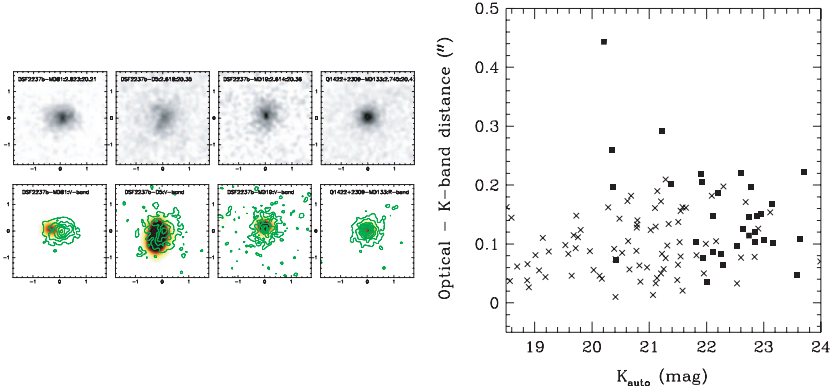}
\end{center}
\caption{
Left) $K$-band images (upper) and optical images with $K$-band contour (lower)
of the 4 brightest $z\sim3$ LBGs
in the sample. Each panel has a $3.\!^{\prime\prime}5 \times 3.\!^{\prime\prime}5$
(27kpc$\times$27kpc at $z=3$) FoV. 
North is to the top and east is to the left. 
We do not match the PSF size of the $K$-band
contour to that of the optical images, thus only the positions of the peaks should
be compared.
Right) Distance between $K$-band and optical peaks as a function of
$K$-band magnitudes for the $z\sim3$ LBGs (filled squares) and
field object (crosses).
}\label{akiyama_f01}
\end{figure}

The AO-assisted $K$-band images of the 4 brightest $z\sim3$ LBGs
are shown in the upper left panel of Figure~\ref{akiyama_f01} as examples. 
The AO observations clearly resolve most of the 31 $z\sim3$ LBGs at
the resolution of FWHM$\sim0.\!^{\prime\prime}2$. 
% The median $r_{\rm HL}$
% is $0.\!^{\prime\prime}23$, which corresponds to 1.8 kpc at $z=3$. 
% LBGs brighter than $M_{V}^{*}$ 
% have larger $r_{\rm HL}$ ($0.\!^{\prime\prime}40$) than the fainter LBGs 
% ($0.\!^{\prime\prime}23$)
% on average, and there is no bright LBGs with small $r_{\rm HL}$.

We examined their total $J-K$ colors, which corresponds to the rest-frame
$U-V$ colors, using seeing-limited $J$-band data of
thirty of the LBGs.
The bright LBGs show red rest-frame $U-V$ colors
(average of $0.2$ mag), while
most of the fainter LBGs show blue rest-frame $U-V$ color (average of $-0.4$ mag).
The color distribution of the LBGs show that the LBG sample covers
not only blue less massive galaxies but also red massive galaxies at among $2<z<3$
galaxies selected with photometric redshifts.

In the lower left panels of Figure~\ref{akiyama_f01}, the
$K$-band contours of the LBGs are compared with the
optical images with the seeing-limited resolution. 
The peaks in the $K$-band images of the LBGs show
significant offset from those in the optical images. In the right
panel of Figure~\ref{akiyama_f01}, the shifts of the LBGs are shown as a 
function of $K$-band magnitude with filled squares for the whole sample. In order to 
evaluate the uncertainties of the shift measurements, we also plot the differences
of $K$-band and optical positions of compact field objects with crosses.
The peaks in the $K$-band images of 7 of the LBGs with $K<22$ mag
show significant or marginal shifts from those in the optical images.
The presence of the shifts among red luminous LBGs implies that the UV-bright star-forming regions
are not necessarily centered at their main body observed in the $K$-band.
% For the LBGs with $K>22$ mag, no shift between the peaks in 
% the $K$-band and the optical images is observed.

\begin{figure}[t]
\begin{center}
\includegraphics[width=110mm]{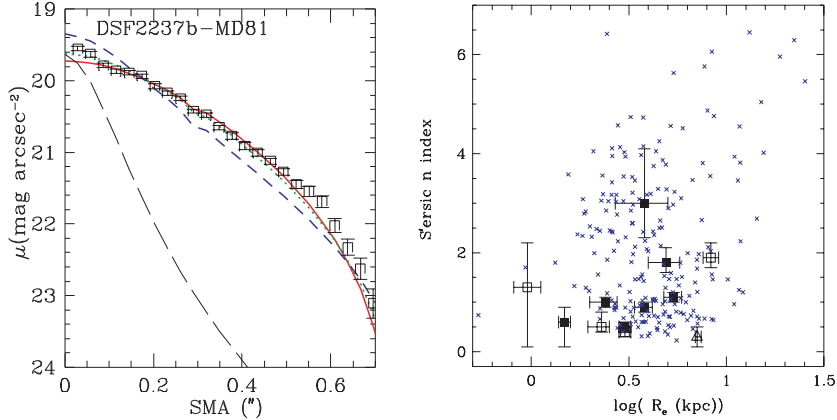}
\end{center}
\caption{
Left) $K$-band profiles of an LBG along the semi-major axis.
The red-solid, blue-dashed, and green-dotted lines show the best
fit S\'ersic (with free $n$), $r^{1/4}$ (S\'ersic with $n=4$), and
exponential (S\'ersic with $n=1$) profiles, respectively. Long dashed
line show the estimated profiles of the PSFs at the object positions.
Right) $R_{e}$ vs. $n$ for the $z\sim3$ LBGs (filled squares) and DRGs (open squares)
with $K<21.5$ mag. Simulated $z=3$ galaxies that are brighter than $K=21.5$ mag
in the 2 mag PLE model are plotted with blue small crosses.}\label{akiyama_f02}
\end{figure}

In order to examine whether their light profiles are similar
to those of local galaxies or not, we apply one component S\'ersic
profile fitting to the AO-assisted $K$-band images of the LBGs with $K<21.5$mag.
An example of the S\'ersic profile fitting is shown in the left panel
of Figure~\ref{akiyama_f02} (DSF2237b-MD81 with $K=20.2$ mag at $z=2.82$). 
As can be seen in the panel, the exponential profile (S\'ersic profile with
$n=1$) fit the profile better than the $r^{1/4}$ profile (with $n=4$), and
the best-fit $n$ is 0.9 with free $n$ fitting. 
The results of the S\'ersic profile fitting of the LBGs 
with $K<21.5$ mag are shown in the right panel of Figure~\ref{akiyama_f02}
with filled and open squares, respectively. We also examined the profiles
of serendipitously observed Distant Red Galaxies in the FoVs.
The images of all but one of the LBGs and DRGs with $K<21.5$ mag are fitted
well with S\'ersic profile with $n$ index less than 2, similar to
disk galaxies in the local universe. 
% We take care of the uncertainties of the estimate of the final PSF at
% the positions of the targets due to the variation of the PSF inside
% the FoV. We check the dependency of the best-fit parameters on the
% PSF shape used in the fitting. The uncertainties shown in the right
% panel of the figure include the effect of the PSF uncertainty.

In order to directly compare the best-fit parameters with those
of galaxies at intermediate redshifts, we make simulated images of
$z\sim3$ galaxies by "cloning" HST/ACS images of galaxies at
$z=0.4-0.7$. Their distribution is shown with blue crosses in the
panel. 
The LBGs and DRGs locate similar part of the $R_{e}$ vs. $n$ plane
to the disk galaxies at intermediate redshifts.

\begin{figure}[t]
\begin{center}
\includegraphics[width=110mm]{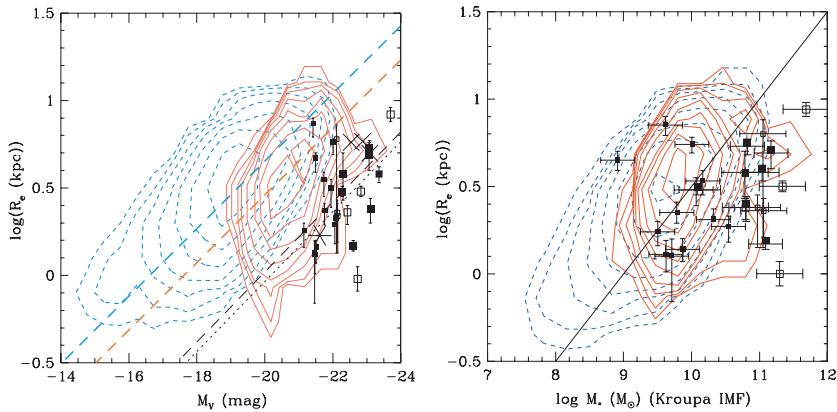}
\end{center}
\caption{Left) $M_{V}$ vs. $R_e$ for the LBG sample (filled squares). Large
and small symbols indicate $R_{e}$ obtained with the S\'ersic profile fitting
with free $n$ and fixed $n=1$, respectively. Open squares show the DRG sample.
Contours show the distributions
of $z=0$ (dashed blue contour) and $z=1$ (solid red contour) disk-galaxies
from Barden et al. (2005). The blue and red dashed lines show $<\mu_{V}>$
of disk galaxies at $z=0$ ($20.84$ mag arcsec$^{-2}$) and $z=1$ (19.84 mag arcsec$^{-2}$),
respectively.
Right) $M_{*}$ vs. $R_{e}$ of the LBG (filled squares) and DRG (open squares)
samples. Contours show the distributions of $z=0$ and $1$ disk-galaxies
from Barden et al. (2005). The solid line represents the relation $\log \Sigma_{M}
(M_{\odot}$kpc$^{-2})=8.50$ derived from the disk galaxies at $z=0-1$
(Barden et al. 2005).}\label{akiyama_f03}
\end{figure}

Assuming that the $z\sim3$ LBGs and DRGs have a disk shape, we 
compare their size-luminosity and size-stellar mass relations with
those of disk galaxies at low/intermediate redshifts in 
the left and right panels of Figure~\ref{akiyama_f03}, respectively.
It should be noted that still there is a possibility that the $z\sim3$ LBGs
and DRGs have a spheroidal shape with $n=1$ profile instead of the disk shape.
The $z\sim3$ LBGs are brighter than $z=0$ and $1$ disk
galaxies at the same $R_{e}$. The surface brightness of the LBGs,
which are estimated from $M_{V}$ and $R_{e}$, are 2.2-2.9 mag
and 1.2-1.9 mag brighter than those of the disk galaxies at $z=0$ and $1$,
respectively. 
The size-stellar mass relation indicates that
the $z\sim3$ LBGs brighter than $M_{V}^{*}$ have
the average
surface stellar mass density 3-6 times larger than those 
of the $z=0$ and $1$ disk galaxies. 
On the contrary, for less-luminous $z\sim3$ LBGs, 
their size-stellar mass relation is similar to
those of $z=0$ and $1$ disk galaxies.

\section{Speculation: LBGs before $z\sim3$ and after}

Because in the local universe, such 
disk with high surface stellar mass density as the massive $z\sim3$ LBGs is rare, we expect that the disks of the
massive $z\sim3$ LBGs are destroyed between $z\sim3$ to $1$. The
strong spatial clustering of the LBGs implies that they reside in massive
dark halos and that they evolve into local spheroids (Adelberger et al. 2005).
Therefore, the disks with high surface stellar mass density would evolve into
local spheroids through "dry" merging events.

Considering the difference between the mass of the dark matter
halos the $z\sim3$ LBGs reside ($2-6\times10^{11} M_{\odot}$; 
Adelberger et al. 2005) and the Jeans mass at the time of
reionization ($\sim10^{10} M_{\odot}$), we naively expect that the
$z\sim3$ LBGs form from builing blocks through several major merges.
The high fraction of disk-like galaxies in the high-redshift universe
would be explained with a hypothesis that they have been going through
only gas-rich "wet" merges at that time (e.g., Springel \& Hernquist 2005). 

We start an AO observing program with laser guide stars in order to
extend the sample of bright $z\sim3$ galaxies, and to
establish the morphological evolution from $z=3$ to $0$.
The high-resolution $K$-band imaging observations of the galaxies at $z\ge3$
should be one of the unique fields that can be explored only with LGS AO on
ground-based 8-10m class telescopes until the launch of the James Webb Space Telescope.

% \begin{acknowledgments}
% Part of the program is supported by a Grant-in-Aid for Young Scientists (B) from
% Japan Society for the Promotion of Science (18740118).
% \end{acknowledgments}

% \begin{discussion}
% \discuss{}{}
% \end{discussion}

\end{document}